\documentclass[twocolumn,amsmath,amssymb,superscriptaddress]{revtex4-1}

\usepackage{graphicx}

\usepackage{amstext,amsmath,amssymb,amsfonts}
\usepackage{bm}

\usepackage{txfonts}
\usepackage{mathtools}
\DeclareMathAlphabet{\mathpzc}{OT1}{pzc}{m}{it}
\DeclareFontFamily{OT1}{pzc}{}
\DeclareFontShape{OT1}{pzc}{m}{it}{<-> s * [1.100] pzcmi7t}{}
\DeclareMathAlphabet{\mathpzc}{OT1}{pzc}{m}{it}

\usepackage{hyperref}

\usepackage{color}
\definecolor{lightblue}{rgb}{0.2,0.2,0.7}
\definecolor{darkblue}{rgb}{0,0.25,0.5}
\definecolor{redbrown}{rgb}{0.875,0.25,0.125}
\definecolor{darkgreen}{rgb}{0,0.5,0}

\renewcommand{\b}[1]{\ensuremath{\mathbf{#1}}}
\renewcommand{\H}{\ensuremath{\text{H}}}

\renewcommand{\l}{\ensuremath{\lambda}}

\newcommand{\lr}{\ensuremath{\text{lr}}}
\newcommand{\sr}{\ensuremath{\text{sr}}}
\newcommand{\ee}{\ensuremath{\text{ee}}}

\newcommand{\HF}{\ensuremath{\text{HF}}}

\renewcommand{\d}{\ensuremath{\text{d}}}
\newcommand{\s}{\ensuremath{\text{s}}}

\newcommand{\x}{\ensuremath{\text{x}}}
\newcommand{\xc}{\ensuremath{\text{xc}}}

\newcommand{\Hxc}{\ensuremath{\text{Hxc}}}

\DeclareMathOperator{\erf}{erf}

\renewcommand{\i}{\ensuremath{\text{i}}}

\begin{document}

\title{Photoionization and core resonances from range-separated time-dependent density-functional theory for open-shell states: Example of the lithium atom}
\author{Julien Toulouse}
\email{toulouse@lct.jussieu.fr}
\affiliation{Laboratoire de Chimie Th\'eorique, Sorbonne Universit\'e and CNRS, F-75005 Paris, France}
\affiliation{Institut Universitaire de France, F-75005 Paris, France}

\author{Karno Schwinn}
\affiliation{Laboratoire de Chimie Th\'eorique, Sorbonne Universit\'e and CNRS, F-75005 Paris, France}
\author{Felipe Zapata}
\affiliation{Department of Physics, Lund University, Box 118, SE-221 00 Lund, Sweden}
\author{Antoine Levitt}
\affiliation{Laboratoire de math\'ematiques d'Orsay, Universit\'e Paris-Saclay and CNRS, F-91405 Orsay, France}
\author{\'Eric Canc\`es}
\affiliation{CERMICS, \'Ecole des Ponts and Inria Paris, 6 \& 8 Avenue Blaise Pascal, 77455 Marne-la-Vall\'ee, France}
\author{Eleonora Luppi}
\email{eleonora.luppi@lct.jussieu.fr}
\affiliation{Laboratoire de Chimie Th\'eorique, Sorbonne Universit\'e and CNRS, F-75005 Paris, France}

\date{November 12, 2022}

\begin{abstract}
We consider the calculations of photoionization spectra and core resonances of open-shell systems using range-separated time-dependent density-functional theory. Specifically, we use the time-dependent range-separated hybrid (TDRSH) scheme, combining a long-range Hartree-Fock (HF) exchange potential and kernel with a short-range potential and kernel from a local density-functional approximation, and the time-dependent locally range-separated hybrid (TDLRSH) scheme, which uses a local range-separation parameter. To efficiently perform the calculations, we formulate a spin-unrestricted linear-response Sternheimer approach in a non-orthogonal B-spline basis set and using appropriate frequency-dependent boundary conditions. We illustrate this approach on the Li atom, which suggests that TDRSH and TDLRSH are adequate simple methods for estimating single-electron photoionization spectra of open-shell systems.
\end{abstract}

\maketitle

\section{Introduction}

Adiabatic linear-response time-dependent density-functional theory (TDDFT)~\cite{RunGro-PRL-84,GroKoh-PRL-85,Cas-INC-95,PetGosGro-PRL-96}, using range-separated approximations~\cite{TawTsuYanYanHir-JCP-04,YanTewHan-CPL-04,PeaHelSalKeaLutTozHan-PCCP-06,LivBae-PCCP-07,BaeLivSal-ARPC-10,TsuSonSuzHir-JCP-10,FroKneJen-JCP-13,RebSavTou-MP-13}, is recognized as a practical and reasonably accurate approach for calculating bound-state electronic excitations in many molecular systems (see, e.g., Refs.~\onlinecite{JacPerScuCioAda-JCTC-08,LauJac-IJQC-13}). A natural question is then whether these range-separated TDDFT methods can be also successfully extended from bound to continuum excitations in order to calculate for example photoionization spectra and resonances in atomic and molecular systems.

In Ref.~\onlinecite{ZapLupTou-JCP-19}, some of the present authors started to explore the merits of range-separated TDDFT for the calculation of photoionization spectra and showed that the so-called linear-response time-dependent range-separated hybrid (TDRSH) scheme~\cite{RebSavTou-MP-13,TouRebGouDobSeaAng-JCP-13}, combining a long-range Hartree-Fock (HF) exchange potential and kernel with a short-range potential and kernel from a local density-functional approximation, provides an adequate (single-electron) photoionization spectrum of the He atom. Technically, in Ref.~\onlinecite{ZapLupTou-JCP-19}, the continuum was described by the use of B-spline basis set within a computational box and the photoionization spectrum was straightforwardly calculated by diagonalization of the linear-response Casida equations (in the orthogonal occupied/virtual orbital basis) using zero boundary conditions at the edge of the box. In Ref.~\onlinecite{SchZapLevCanLupTou-JCP-22}, the present authors extended this study to the Be atom and showed that the TDRSH scheme and a close variant, namely the time-dependent locally range-separated hybrid (TDLRSH) scheme, also give overall reasonable (single-electron) photoionization spectra for this system, with core resonances at approximately correct resonance energies, albeit with much too small resonance widths. To be able to efficiently apply TDRSH and TDLRSH to the Be atom, in Ref.~\onlinecite{SchZapLevCanLupTou-JCP-22}, we use a linear-response Sternheimer approach~\cite{MahSub-BOOK-90,SteDecLis-JPB-95,SteDec-JCP-00,YabNakIwaBer-PSS-06,AndBotMarRub-JCP-07,StrLehRubMarLou-INC-12,HofSchKum-JCP-18,HofSchKum-PRA-19,HofKum-JCP-20} (in the non-orthogonal B-spline basis) using appropriate frequency-dependent boundary conditions at the edge of the computational box. 

The work of Ref.~\onlinecite{SchZapLevCanLupTou-JCP-22} was restricted to systems with closed-shell ground-state states. In the present work, we extend the theory of Ref.~\onlinecite{SchZapLevCanLupTou-JCP-22} to systems with open-shell ground-state states. For this, we provide equations for a general linear-response Sternheimer scheme (in a non-orthogonal basis set) within a spin-unrestricted formalism, and again with frequency-dependent boundary conditions at the edge of the computational box. As an illustration, we use this scheme to calculate photoionization spectra of the Li atom at the TDRSH and TDLRSH levels, we extract Fano parameters of some of the core resonances, and we compare with the standard time-dependent local-density approximation (TDLDA) and time-dependent Hartree-Fock (TDHF) methods. 

The paper is organized as follows. In Section~\ref{sec:theory}, we review the range-separated hybrid (RSH) and locally range-separated hybrid (LRSH) schemes in a spin-unrestricted formalism, and give in some detail the linear-response spin-unrestricted Sternheimer equations including a nonlocal HF exchange kernel both in real space and in a general non-orthogonal basis set which, to the best of our knowledge, were never given in the literature. We also give some computational details for our specific implementation for the Li atom using a B-spline basis set. In Section~\ref{sec:results}, we give and discuss the results obtained on the Li atom. We explain how to select an optimal range-separation parameter, we discuss the calculated photoionization spectra, and we analyze the core resonances. Section~\ref{sec:conclusion} contains our conclusions.

\section{Theory and computational method}
\label{sec:theory}

We work on the one-electron Hilbert space $L^2(\mathbb{R}^3_{\Sigma},\mathbb{C})$ where $\mathbb{R}^3_{\Sigma} = \mathbb{R}^3 \times \Sigma$ and $\Sigma=\{\uparrow,\downarrow\}$ is the set of spin coordinates. We denote a space-spin electron coordinate as $\b{x}=(\b{r},s) \in \mathbb{R}^3_{\Sigma}$. We use throughout a spin-unrestricted formalism. Unless otherwise indicated, Hartree atomic units are used in this work.

\subsection{Range-separated hybrid scheme}

In the range-separated hybrid (RSH) scheme~\cite{AngGerSavTou-PRA-05}, the spin-orbitals $\{\varphi_i\}$ and their associated energies $\{\varepsilon_i\}$ of an $N$-electron system are found from the self-consistent Schr\"odinger-type equation
\begin{eqnarray}
\int_{\mathbb{R}^3_\Sigma} h[\gamma_0](\b{x},\b{x}') \varphi_i(\b{x}') \d \b{x}' = \varepsilon_i \varphi_i(\b{x}),
\label{}
\end{eqnarray}
where $h[\gamma_0](\b{x},\b{x}')$ is the nonlocal RSH Hamiltonian depending on the density matrix $\gamma_0(\b{x},\b{x}') = \sum_{i=1}^{N} \varphi_i(\b{x}) \varphi_i^*(\b{x}')$. The RSH Hamiltonian has the form, for a generic density matrix $\gamma$,
\begin{eqnarray}
h[\gamma](\b{x},\b{x}') = T(\b{x},\b{x}') + \delta(\b{x}-\b{x}') v_{\text{ne}}(\b{r}) + v_\Hxc[\gamma](\b{x},\b{x}'), \;
\label{}
\end{eqnarray}
where $T(\b{x},\b{x}')$ is the kinetic integral kernel such that $\int_{\mathbb{R}^3_\Sigma} T(\b{x},\b{x}') \varphi_i(\b{x}') \d \b{x}' = -(1/2) \bm{\nabla}^2_{\b{r}} \varphi_i(\b{x})$, and $v_{\text{ne}}(\b{r})$ is the nuclei-electron potential and $v_\Hxc[\gamma](\b{x},\b{x}')$ is the Hartree-exchange-correlation potential. The expression of $v_\Hxc[\gamma](\b{x},\b{x}')$ is
\begin{eqnarray}
v_\Hxc[\gamma](\b{x},\b{x}') &=& \delta(\b{x}-\b{x}') v_\H[\rho_\gamma](\b{r}) + v_\x^{\lr,\HF}[\gamma](\b{x},\b{x}') 
\nonumber\\
&& + \delta(\b{x}-\b{x}') v_\xc^{\sr}[\rho_{\gamma}](\b{x}),
\label{vHxc}
\end{eqnarray}
containing the local Hartree potential 
\begin{eqnarray}
v_\text{H}[\rho_\gamma](\b{r}) = \int_{\mathbb{R}^3_\Sigma} \rho_\gamma(\b{x}') w_\ee(\b{r},\b{r}') \d \b{x}',
\end{eqnarray}
written with the spin-resolved density $\rho_\gamma(\b{x})=\gamma(\b{x},\b{x})$ and the Coulomb electron-electron interaction $w_\ee(\b{r},\b{r}')=1/|\b{r}-\b{r}'|$, the nonlocal long-range (lr) HF exchange potential 
\begin{eqnarray}
v_{\text{x}}^{\lr,\HF}[\gamma](\b{x},\b{x}') = - \gamma(\b{x},\b{x}') w_\ee^\lr(\b{r},\b{r}'),
\label{vxlrHF}
\end{eqnarray}
written with the long-range electron-electron interaction~\cite{Sav-INC-96} 
\begin{eqnarray}
w_\ee^\lr(\b{r},\b{r}')=\frac{\erf(\mu|\b{r}-\b{r}'|)}{|\b{r}-\b{r}'|},
\label{weelrerf}
\end{eqnarray}
with $\mu = \tilde{\mu}/a_0$ where $a_0=1$ a.u. is the Bohr radius and $\tilde{\mu}\in [0,+\infty)$ is the adimensional range-separation parameter, and the local complementary short-range (sr) exchange-correlation potential $v_\xc^{\sr}[\rho_\gamma](\b{x})$. For the latter term, we use in this work the LDA
\begin{eqnarray}
v_\xc^{\sr}[\rho_\gamma](\b{r},s) = \frac{\partial \bar{e}_{\xc,\text{UEG}}^\sr(\rho_\uparrow,\rho_\downarrow,\mu)}{\partial \rho_s} \Biggl|_{\substack{\rho_\uparrow=\rho_\gamma(\b{r},\uparrow)\\\rho_\downarrow=\rho_\gamma(\b{r},\downarrow)}} \;\;,
\label{vxcsr}
\end{eqnarray}
where $\bar{e}_{\xc,\text{UEG}}^\sr(\rho_\uparrow,\rho_\downarrow,\mu)$ is the spin-dependent complementary short-range exchange-correlation energy density of the uniform-electron gas (UEG), as parametrized in Ref.~\onlinecite{PazMorGorBac-PRB-06}.

In the locally range-separated hybrid (LRSH) scheme~\onlinecite{KruScuPerSav-JCP-08,AscKum-JCP-19,KlaBah-JCTC-20,MaiIkaNak-JCP-21,BruBahKum-JCP-22,SchZapLevCanLupTou-JCP-22}, the range-separation parameter $\mu$ in Eqs.~(\ref{weelrerf}) and~(\ref{vxcsr}) is replaced by a function of position $\b{r} \mapsto \mu(\b{r})$. The long-range electron-electron interaction in Eq.~(\ref{weelrerf}) now becomes~\cite{KlaBah-JCTC-20}
\begin{eqnarray}
w_\ee^\lr(\b{r},\b{r}')=\frac{1}{2} \left[ \frac{\erf(\mu(\b{r})|\b{r}-\b{r}'|)}{|\b{r}-\b{r}'|} + \frac{\erf(\mu(\b{r}')|\b{r}-\b{r}'|)}{|\b{r}-\b{r}'|}\right].
\label{weelrerfmur}
\end{eqnarray}
Following Ref.~\onlinecite{KruScuPerSav-JCP-08}, we choose $\mu(\b{r})$ as
\begin{eqnarray}
\mu(\b{r}) = \frac{\tilde{\mu}}{2} \frac{|\nabla \rho(\b{r})|}{\rho(\b{r})},
\end{eqnarray}
where again $\tilde{\mu} \in [0,+\infty)$ is the adimensional range-separation parameter and we take $\rho(\b{r})$ as the fixed spin-unrestricted Hartree-Fock (UHF) ground-state density.

\subsection{Linear-response Sternheimer equations in real space}

We consider a time-dependent perturbation potential of the form
\begin{eqnarray}
v_\text{ext}(\b{r},t) = \left[ v_\text{ext}(\b{r}) e^{-\i \omega t} + v_\text{ext}(\b{r}) e^{+\i \omega t} \right] e^{\eta t},
\label{vextt}
\end{eqnarray}
where $v_\text{ext}(\b{r}) = \b{r} \, \cdot \, {\cal E} \, \b{e}$ is the electric-dipole interaction (${\cal E}$ is the amplitude of the electric field and $\b{e}$ is its unit polarization vector), $\omega \geq 0$ is the frequency, and $e^{\eta t}$ is an adiabatic switching factor with a small parameter $\eta > 0$. Following the same steps as in Ref.~\onlinecite{SchZapLevCanLupTou-JCP-22}, it can be shown that the Fourier components at frequencies $\pm \omega +\i \eta$ of the first-order change of the occupied spin-orbital $\varphi_i$, namely $\psi_i^{(+)}$ and $\psi_i^{(-)}$, are given by the following TDRSH or TDLRSH Sternheimer equations
\begin{widetext}
\begin{eqnarray}
\!\! \left(\pm \omega +\i \eta +\varepsilon_i \right) \psi_i^{(\pm)}(\b{x}_1,\omega) &=& \! \! \int_{\mathbb{R}^3_\Sigma} \! h[\gamma_0](\b{x}_1,\b{x}_1') \psi_i^{(\pm)}(\b{x}_1',\omega) \d\b{x}_1'
\nonumber\\
&& + \!\! \int_{\mathbb{R}^9_\Sigma} f_\Hxc[\gamma_0](\b{x}_1,\b{x}_1';\b{x}_2,\b{x}_2') \gamma^{(\pm)}(\b{x}_2,\b{x}_2',\omega) \varphi_i(\b{x}_1') \d\b{x}_1' \d\b{x}_2 \d\b{x}_2'
+v_\text{ext}^{(1)}(\b{r}_1) \varphi_i(\b{x}_1), \;\;
\label{Sternheimerrealspace}
\end{eqnarray}
\end{widetext}
with the first-order perturbation potential $v_\text{ext}^{(1)}(\b{r}) = \b{r} \cdot \b{e}$, the first-order changes of the density matrix
\begin{eqnarray}
\gamma^{(\pm)}(\b{x},\b{x}',\omega) = \sum_{i=1}^{N} \left[\psi_i^{(\pm)}(\b{x},\omega) \varphi_i^*(\b{x}') + \varphi_i(\b{x}) \psi_i^{(\mp)*}(\b{x}',\omega) \right],
\nonumber\\
\label{gammapm}
\end{eqnarray}
and the Hartree-exchange-correlation kernel
\begin{eqnarray}
f_\Hxc[\gamma_0](\b{x}_1,\b{x}_1';\b{x}_2,\b{x}_2') = \delta(\b{x}_1-\b{x}_1') \delta(\b{x}_2-\b{x}_2') f_\H(\b{r}_1,\b{r}_2) \;\;\;\;
\nonumber\\
+ f_\H^{\lr,\HF}(\b{x}_1,\b{x}_1';\b{x}_2,\b{x}_2')
+ \delta(\b{x}_1-\b{x}_1') \delta(\b{x}_2-\b{x}_2') f_\xc^{\sr}[\rho_{\gamma_0}](\b{x}_1,\b{x}_2),
\nonumber\\
\label{}
\end{eqnarray}
where $f_\H(\b{r}_1,\b{r}_2) = w_\ee(\b{r}_1,\b{r}_2)$ is the Hartree kernel, $f_\H^{\lr,\HF}(\b{x}_1,\b{x}_1';\b{x}_2,\b{x}_2') = -\delta(\b{x}_1-\b{x}_2) \delta(\b{x}_1'-\b{x}_2') w_\ee^{\lr}(\b{r}_1,\b{r}_1')$ is the nonlocal HF exchange kernel, and $f_\xc^{\sr}[\rho_{\gamma_0}](\b{x}_1,\b{x}_2)$ is the short-range exchange-correlation kernel, which for the LDA [Eq.~(\ref{vxcsr})] takes the local form
\begin{eqnarray}
f_\xc^{\sr}[\rho_{\gamma_0}](\b{r}_1,s_1,\b{r}_2,s_2) = \phantom{xxxxxxxxxxxxxxxxxxxxxx}
\nonumber\\
 \delta(\b{r}_1-\b{r}_2)  \left. \frac{\partial^2 \bar{e}_{\xc,\text{UEG}}^\sr(\rho_\uparrow,\rho_\downarrow,\mu)}{\partial \rho_{s_1}\partial \rho_{s_2}} \right|_{\substack{\rho_\uparrow=\rho_{\gamma_0}(\b{r}_1,\uparrow)\\\rho_\downarrow=\rho_{\gamma_0}(\b{r}_1,\downarrow)}}.
\label{fxcsr}
\end{eqnarray}

The photoexcitation/photoionization cross section can then be calculated as~\cite{HofSchKum-JCP-18}
\begin{eqnarray}
\sigma(\omega) = \lim_{\eta \to 0^+} \frac{4\pi \omega}{c} \text{Im}[\alpha(\omega + \i \eta)],
\label{sigma}
\end{eqnarray}
where $c = 137.036$ a.u. is the speed of light and $\alpha(\omega)$ is the spherically averaged dipole polarizability given by
\begin{eqnarray}
\alpha(\omega + \i \eta) = - \frac{1}{3} \sum_{a\in \{x,y,z\}} \int_{\mathbb{R}^3_\Sigma} (\b{r} \cdot \b{u}_a) \; \rho^{(+)} (\b{x},\omega) \d\b{x},
\label{alpha}
\end{eqnarray}
where $\b{u}_a$ is the unit vector along the direction $a$ and $\rho^{(+)}(\b{x},\omega) = \gamma^{(+)} (\b{x},\b{x},\omega)$.

\vspace{0.5cm}
\subsection{Linear-response Sternheimer equations in a basis set}
\label{sec:sternheimer_basis_set}

Let us introduce now a finite (non-orthogonal) spatial basis set on a domain $\Omega \subset \mathbb{R}^3$, i.e. 
$\{\chi_\nu\}_{\nu=1,...,M} \subset H^1(\Omega,\mathbb{C})$ (where $H^1$ is the first-order Sobolev space), made of
$M$ basis functions to expand the occupied spin-orbitals
\begin{eqnarray}
\varphi_j(\b{r},s) = \delta_{s_j,s} \sum_{\nu=1}^{M} c_{j\nu} \chi_\nu(\b{r}),
\label{phijexpand}
\end{eqnarray}
where $s_j\in \Sigma$ is the spin of the spin-orbital $j$, and their first-order changes
\begin{eqnarray}
\psi_j^{(\pm)}(\b{r},s,\omega) = \delta_{s_j,s} \sum_{\nu=1}^{M} c_{j\nu}^{(\pm)}(\omega) \chi_\nu(\b{r}),
\label{psijexpand}
\end{eqnarray}
where $c_{j\nu}$ and $c_{j\nu}^{(\pm)}(\omega)$ are (generally complex-valued) coefficients labeled with the composite index $j\nu \equiv (j,\nu) \in  \llbracket  1, N\rrbracket \times \llbracket  1, M \rrbracket$. Integrating Eq.~(\ref{Sternheimerrealspace}) against a basis function $\chi_\mu^*$, and using the expansions of Eqs.~(\ref{phijexpand}) and~(\ref{psijexpand}), 
leads to the basis-set Sternheimer equations in the following block matrix form
\begin{eqnarray}
\left( \begin{array}{cc}
\bm{\Lambda}(\omega) & \b{B} \\
\b{B}^* & \b{\Lambda}(-\omega)^*
\end{array} \right)
\left( \begin{array}{c}
\b{c}^{(+)}(\omega)  \\
\;\b{c}^{(-)}(\omega)^* \\
\end{array} \right)
=-\left( \begin{array}{c}
\b{V}  \\
\;\b{V}^* \\
\end{array} \right),
\label{Sternheimermatrix}
\end{eqnarray}
which must be solved at each given frequency $\omega$ for $\b{c}^{(+)}(\omega)$ and $\b{c}^{(-)}(\omega)^*$ which are the column vectors of components $c^{(+)}_{j\nu}(\omega)$ and $c^{(-)}_{j\nu}(\omega)^*$, respectively. In Eq.~(\ref{Sternheimermatrix}), $\b{V}$ is the column vector of components $V_{i\mu}= \b{e} \cdot \sum_{\nu=1}^M \b{d}_{\mu,\nu} c_{i\nu}$ where $\b{d}_{\mu,\nu}= \int_{\Omega} \chi_\mu^*(\b{r}) \b{r} \chi_\nu(\b{r}) \d\b{r}$ are the dipole-moment integrals, and $\bm{\Lambda}(\pm\omega)$ and $\b{B}$ are square matrices with elements
\begin{eqnarray}
\Lambda_{i\mu,j\nu}(\pm\omega) = \delta_{i,j} \left( h_{i,\mu,\nu}(\pm \omega) - (\varepsilon_{i} \pm \omega +\i \eta) S_{\mu,\nu} \right) 
\nonumber\\
+\sum_{\l=1}^{M} \sum_{\sigma=1}^{M} 
\; c_{i\sigma} c_{j\l}^* F_{\mu,\l,\sigma,\nu}^{s_i,s_j},
\label{Lambda}
\end{eqnarray}
and
\begin{eqnarray}
B_{i\mu,j\nu} &=& \sum_{\l=1}^{M} \sum_{\sigma=1}^{M} c_{i\sigma} c_{j\l} F_{\mu,\nu,\sigma,\l}^{s_i,s_j}.
\label{B}
\end{eqnarray}
In Eq.~(\ref{Lambda}), $S_{\mu,\nu} = \int_{\Omega} \chi_\mu^*(\b{r}) \chi_\nu(\b{r}) \d\b{r}$ are the overlap integrals over the basis functions, and $h_{i,\mu,\nu}(\pm \omega)$ are the matrix elements of the RSH or LRSH Hamiltonian
\begin{eqnarray}
h_{i,\mu,\nu}(\pm \omega)       &=& t_{i,\mu,\nu}(\pm \omega) + v_{\mu,\nu}
\\
 &&+ \sum_{\l=1}^{M} \sum_{\sigma=1}^{M} \left( P_{\sigma,\l} w_{\mu,\l,\nu,\sigma} - P_{\sigma,\l}^{s_i} w_{\mu,\l,\sigma,\nu}^\lr \right) + v_{\mu,\nu}^{\sr,s_i},\nonumber
\label{hmunu}
\end{eqnarray}
where $t_{i,\mu,\nu}(\pm \omega)$ are the kinetic integrals
\begin{eqnarray}
  \label{eq:kinetic_integrals}
  t_{i,\mu,\nu}(\pm \omega) &=& \frac 1 2 \int_{\Omega} \nabla \chi_\mu^*(\b{r})\cdot \nabla \chi_\nu(\b{r}) \d\b{r} \\
  &&- \frac{1}{2} \int_{\partial\Omega^{2}} \chi_{\mu}^*(\b r) K_i(\b r,\b r';\pm \omega) \chi_{\nu}(\b r')\d\b{r} \d\b{r'} ,\nonumber
\end{eqnarray}
where $K_i(\b r,\b r';\pm \omega)$ is the Dirichlet-to-Neumann kernel imposing Robin boundary conditions on the surface $\partial \Omega$~\cite{SchZapLevCanLupTou-JCP-22}, $v_{\mu,\nu} = \int_{\Omega} \chi_\mu^*(\b{r}) v_\text{ne}(\b{r}) \chi_\nu(\b{r}) \d\b{r}$ are the nuclei-electron integrals, $P_{\sigma,\l} = \sum_{j=1}^{N} c_{j\sigma} c_{j\l}^*$ are the elements of the total density matrix, $P_{\sigma,\l}^{s_i} = \sum_{j=1}^{N} \delta_{s_j,s_i} c_{j\sigma} c_{j\l}^*$ are the elements of the density matrix of spin $s_i$, $w_{\mu,\l,\nu,\sigma} = \int_{\Omega^2} \chi_\mu^*(\b{r}_1) \chi_\l^*(\b{r}_2) w_\ee(\b{r}_1,\b{r}_2) \chi_\nu(\b{r}_1) \chi_\sigma(\b{r}_2) \d\b{r}_1 \d\b{r}_2$ and $w_{\mu,\l,\sigma,\nu}^\lr = \int_{\Omega^2} \chi_\mu^*(\b{r}_1) \chi_\l^*(\b{r}_2) w_\ee^\lr(\b{r}_1,\b{r}_2) \chi_\sigma(\b{r}_1) \chi_\nu(\b{r}_2) \d\b{r}_1 \d\b{r}_2$ are the Coulombic and long-range two-electron integrals, respectively, and $v_{\mu,\nu}^{\sr,s_i}  = \int_{\Omega} \chi_\mu^*(\b{r}) v_\xc^\sr[\rho_{\gamma_0}](\b{r},s_i) \chi_\nu(\b{r}) \d\b{r}$ are the short-range exchange-correlation potential integrals. In Eqs.~(\ref{Lambda}) and~(\ref{B}), $F_{\mu,\l,\sigma,\nu}^{s_i,s_j}$ comes from the matrix elements of the Hartree-exchange-correlation kernel $f_\Hxc[\gamma_0]$,
\begin{eqnarray}
F_{\mu,\l,\sigma,\nu}^{s_i,s_j} &=& 
 w_{\mu,\l,\sigma,\nu} - \delta_{s_i,s_j} w_{\mu,\l,\nu,\sigma}^\lr + f^{\sr,s_i,s_j}_{\mu,\l,\sigma,\nu},
\end{eqnarray}
where $f^{\sr,s_i,s_j}_{\mu,\l,\sigma,\nu}$ are the short-range exchange-correlation kernel integrals
\begin{eqnarray}
f^{\sr,s_i,s_j}_{\mu,\l,\sigma,\nu} = \phantom{xxxxxxxxxxxxxxxxxxxxxxxxxxxxxxxxxxxxxxxx}
\nonumber\\ 
\int_{\Omega^2} \chi_\mu^*(\b{r}_1) \chi_\l^*(\b{r}_2) f_\xc^\sr[\rho_{\gamma_0}](\b{r}_1,s_i,\b{r}_2,s_j) \chi_\sigma(\b{r}_1) \chi_\nu(\b{r}_2) \d\b{r}_1 \d\b{r}_2 .
\nonumber\\ 
\end{eqnarray}

Finally, in the basis set, the dipole polarizability takes the form
\begin{eqnarray}
\alpha(\omega + \i \eta) =  \phantom{xxxxxxxxxxxxxxxxxxxxxxxxxxxxxxxxx}
\nonumber\\
- \frac{1}{3} \sum_{a\in \{x,y,z\}} \sum_{\mu=1}^M \sum_{\nu=1}^M \left( P_{\mu,\nu}^{(+)}(\omega) \b{d}_{\nu,\mu} + P_{\mu,\nu}^{(-)}(\omega)^* \b{d}_{\nu,\mu}^* \right) \cdot \b{u}_a,
\end{eqnarray}
where $P_{\mu,\nu}^{(\pm)}(\omega)=\sum_{i=1}^{N} c_{i\mu}^{(\pm)}(\omega) c_{i\nu}^*$.

\subsection{Computational details}

We apply the present theory to the Li atom ($N=3$) in the ground-state configuration 1s$^2$2s. We use a dipole interaction with a $z$-polarized electric field, i.e. $v_\text{ext}^{(1)}= \b{r} \cdot \b{u}_z$. The occupied spin-orbitals are of symmetry s ($\ell_i=0$, $m_i=0$) and the perturbed spin-orbitals are of symmetry p$_z$ ($\ell=1$, $m=0$). 

Just like in Ref.~\onlinecite{SchZapLevCanLupTou-JCP-22}, we expand the radial parts of orbitals in a basis set of $M_\s=50$ B-spline functions~\cite{Boo-BOOK-78,BacCorDecHanMar-RPP-01} of order $k_\s=8$, using a constant spatial grid spacing and a maximal radius of $r_{\text{max}} = 25$ bohr. The Robin boundary-condition term in Eq.~(\ref{eq:kinetic_integrals}) takes a simple radial local form, identical to the one used for the Be atom in Ref.~\onlinecite{SchZapLevCanLupTou-JCP-22}. We use $\eta=0$ to avoid artificially broadening of the resonances.

\section{Results and discussion}
\label{sec:results}

We now show and discuss the results on the Li atom.

\subsection{Orbital energies}
\label{sec:orbenergies}

\begin{figure*}
\includegraphics[scale=0.35,angle=-90]{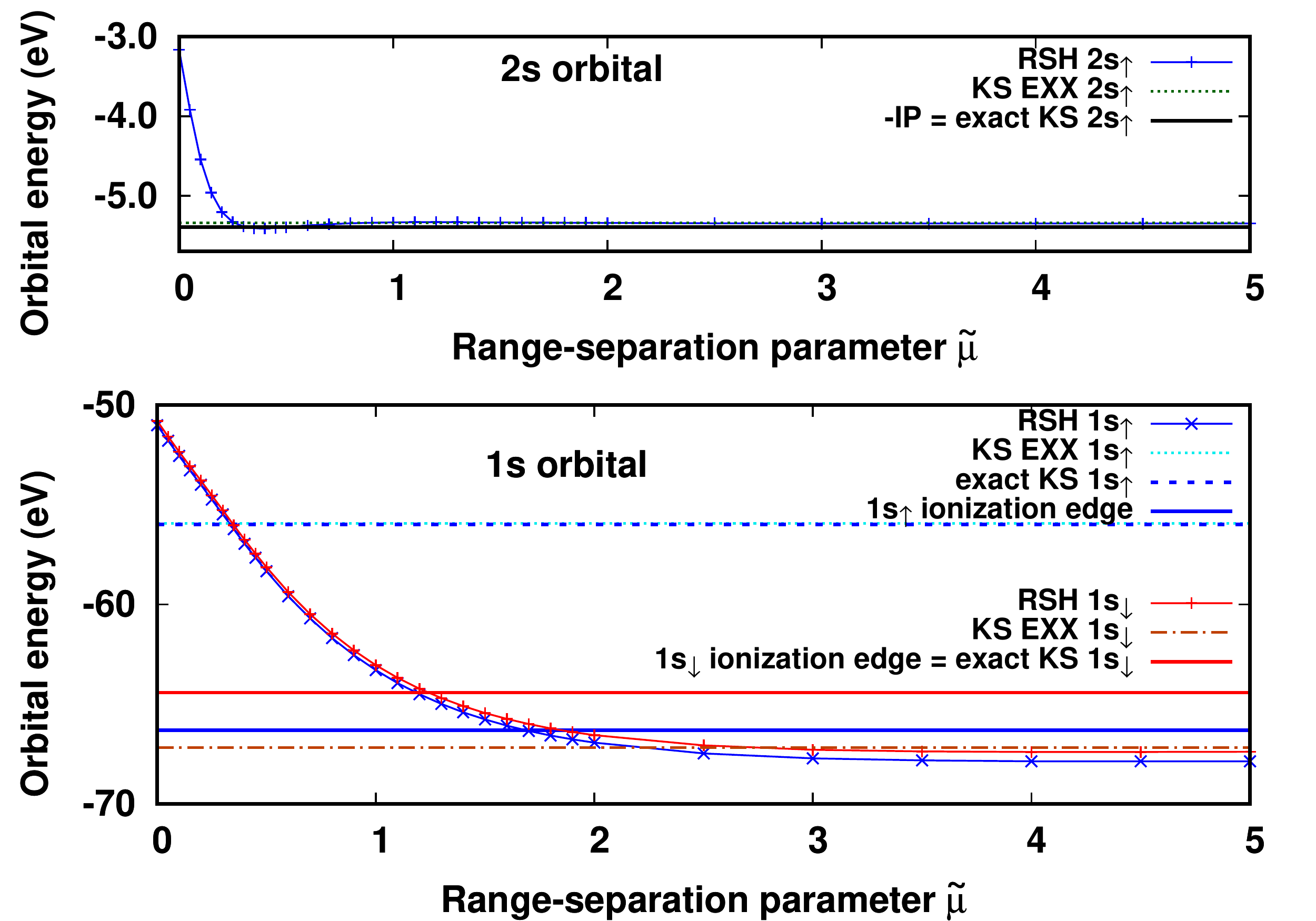}
\includegraphics[scale=0.35,angle=-90]{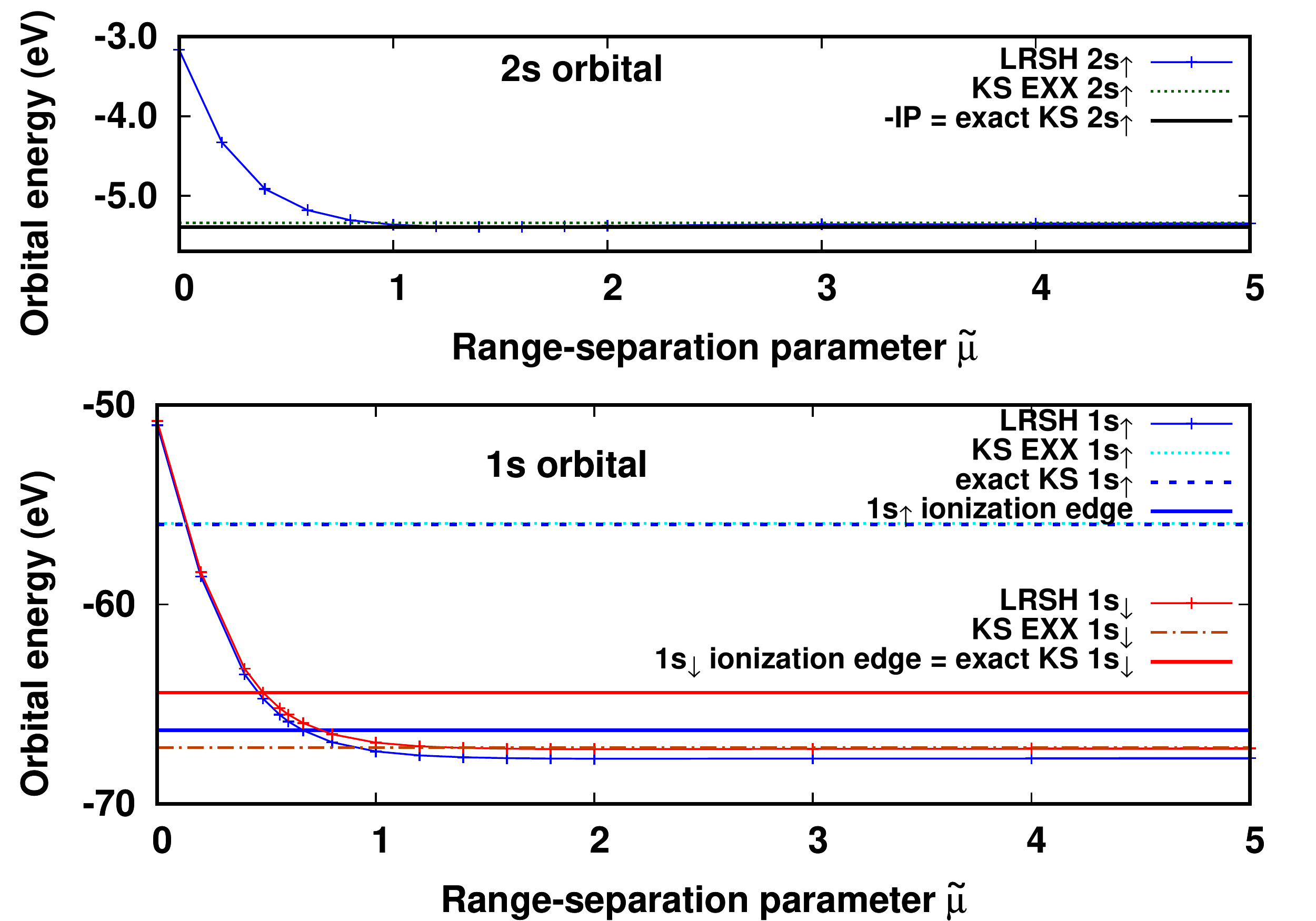}
\caption{RSH and LRSH 1s$_\uparrow$, 1s$_\downarrow$ and 2s$_\uparrow$ spin-orbital energies of the Li atom as a function of the adimensional range-separation parameter $\tilde{\mu}$. As references, the opposite of the experimental IP (-5.392 eV)~\cite{NIST-INC-21} and of the 1s$_\uparrow$ ionization edge (corresponding to the state 1s2s $^1\text{S}$, -66.31 eV)~\cite{LanVieHemMenWehBec-PRA-91} and 1s$_\downarrow$ ionization edge (corresponding to the state 1s2s $^3\text{S}$, -64.41 eV)~\cite{LanVieHemMenWehBec-PRA-91} are indicated, as well as the KS EXX 1s$_\uparrow$, 1s$_\downarrow$, and 2s$_\uparrow$ spin-orbital energies (-55.94 eV, -67.18 eV, and -5.342 eV, respectively)~\cite{CheKriEsqStoIaf-PRA-96} and the exact KS 1s$_\uparrow$ orbital energy (-55.97 eV~\cite{CheKriEsqStoIaf-PRA-96}).}
\label{fig:orbitalenergies}
\end{figure*}

Figure~\ref{fig:orbitalenergies} shows the RSH and LRSH 1s$_\uparrow$, 1s$_\downarrow$ and 2s$_\uparrow$ spin-orbital energies as a function of the adimensional range-separation parameter $\tilde{\mu}$. Also indicated are the opposite of the experimental IP (-5.392 eV)~\cite{NIST-INC-21} and of the 1s$_\uparrow$ ionization edge (corresponding to the two-electron state 1s2s $^1\text{S}$, -66.31 eV)~\cite{LanVieHemMenWehBec-PRA-91} and 1s$_\downarrow$ ionization edge (corresponding to the two-electron state 1s2s $^3\text{S}$, -64.41 eV)~\cite{LanVieHemMenWehBec-PRA-91}, as well as the Kohn-Sham (KS) exact exchange (EXX) 1s$_\uparrow$, 1s$_\downarrow$, and 2s$_\uparrow$ spin-orbital energies (-55.94 eV, -67.18 eV, and -5.342 eV, respectively)~\cite{CheKriEsqStoIaf-PRA-96} and the exact KS 1s$_\uparrow$ spin-orbital energy (-55.97 eV~\cite{CheKriEsqStoIaf-PRA-96}). According to spin-unrestricted KS theory~\cite{AlmBar-PRB-85,GriBae-JCP-02}, the exact KS 2s$_\uparrow$ spin-orbital energy must be equal to the opposite of the exact IP and the exact KS 1s$_\downarrow$ spin-orbital energy must be equal to the opposite of the exact 1s$_\downarrow$ ionization edge. As for the He atom~\cite{ZapLupTou-JCP-19}, the KS EXX spin-orbital energies are rather close to the exact KS spin-orbital energies.

Let us first discuss the 2s$_\uparrow$ spin-orbital energy. At $\tilde{\mu}=0$, both RSH and LRSH reduce to standard KS, and give a far too high 2s$_\uparrow$ spin-orbital energy (by more than 2 eV) due to the well-known self-interaction error of LDA. For $\tilde{\mu}\to \infty$, both RSH and LRSH reduce to standard HF, which gives a 2s$_\uparrow$ spin-orbital energy very close to the opposite of the exact IP (error of only about 0.04 eV). Starting from $\tilde{\mu}=0$, increasing $\tilde{\mu}$ reduces the self-interaction error in the short-range LDA exchange-correlation functional, and the 2s$_\uparrow$ spin-orbital energy essentially reaches its HF value at around $\tilde{\mu} \approx 0.25$ for RSH and $\tilde{\mu} \approx 1$ for LRSH.

Let us now focus on the 1s$_\uparrow$ and 1s$_\downarrow$ spin-orbital energies. Both RSH and LRSH give very small energy splittings between the 1s$_\uparrow$ and 1s$_\downarrow$ spin-orbitals (at most about 0.5 eV for large $\tilde{\mu}$), in comparison with the energy splitting obtained with the exact KS (8.4 eV) and with the experimental ionization edges (1.9 eV). Again, at $\tilde{\mu}=0$, RSH and LRSH reduce to standard KS, and give way too high 1s$_\uparrow$ and 1s$_\downarrow$ spin-orbital energies due to the use of the LDA. For $\tilde{\mu}\to \infty$, when RSH and LRSH reduce to standard HF, we obtain 1s$_\uparrow$ and 1s$_\downarrow$ spin-orbital energies that are too low compared to the experimental ionization edges by about 1.6 and 3.0 eV, respectively. With the present approximations for the short-range exchange-correlation potential and kernel, the RSH and LRSH ionization energies correspond to the opposite of the occupied spin-orbital energies, and are identical to the TDRSH and TDLRSH ionization energies, respectively. In the philosophy of the so-called optimally tuned range-separated hybrids~\cite{LivBae-PCCP-07,SteKroBae-JACS-09,SteKroBae-JCP-09}, in order to obtain correct ionization energies in TDRSH or TDLRSH, it is thus appropriate to choose the optimal adimensional range-separation parameter $\tilde{\mu}$ so that the RSH or LRSH spin-orbital energies are as close as possible to the opposite of the experimental ionization energies. Concretely, since we focus in the work on core excitations, we choose the optimal $\tilde{\mu}$ so as to symmetrically minimize the error in the 1s$_\uparrow$ spin-orbital energy and the error in the 1s$_\downarrow$ spin-orbital energy. This gives optimal adimensional range-separation parameters of $\tilde{\mu}_\text{RSH}=1.431$ for RSH and $\tilde{\mu}_\text{LRSH}=0.560$ for LRSH.

\subsection{Photoionization spectrum}

\begin{figure*}
\includegraphics[scale=0.3,angle=-90]{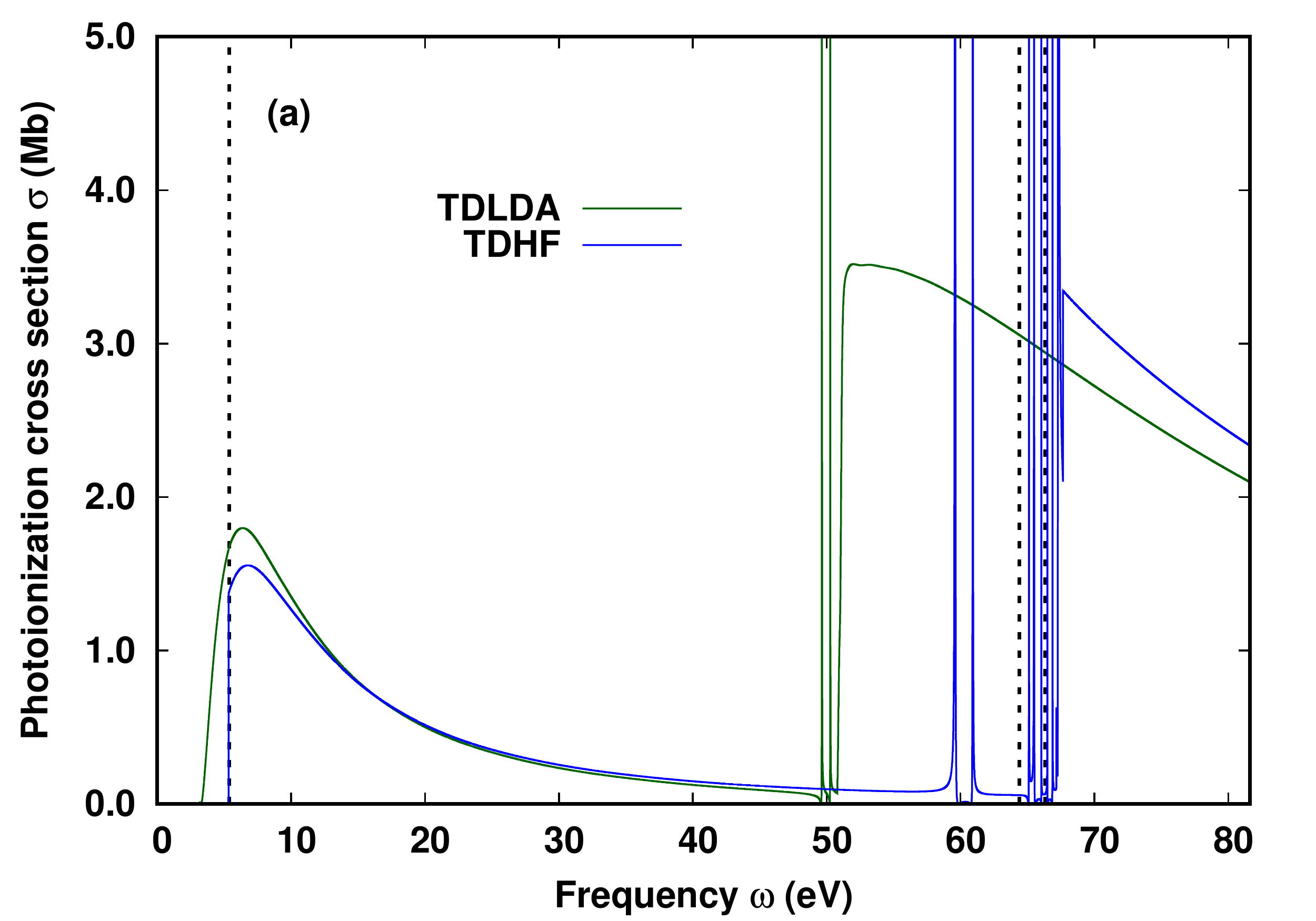}
\includegraphics[scale=0.3,angle=-90]{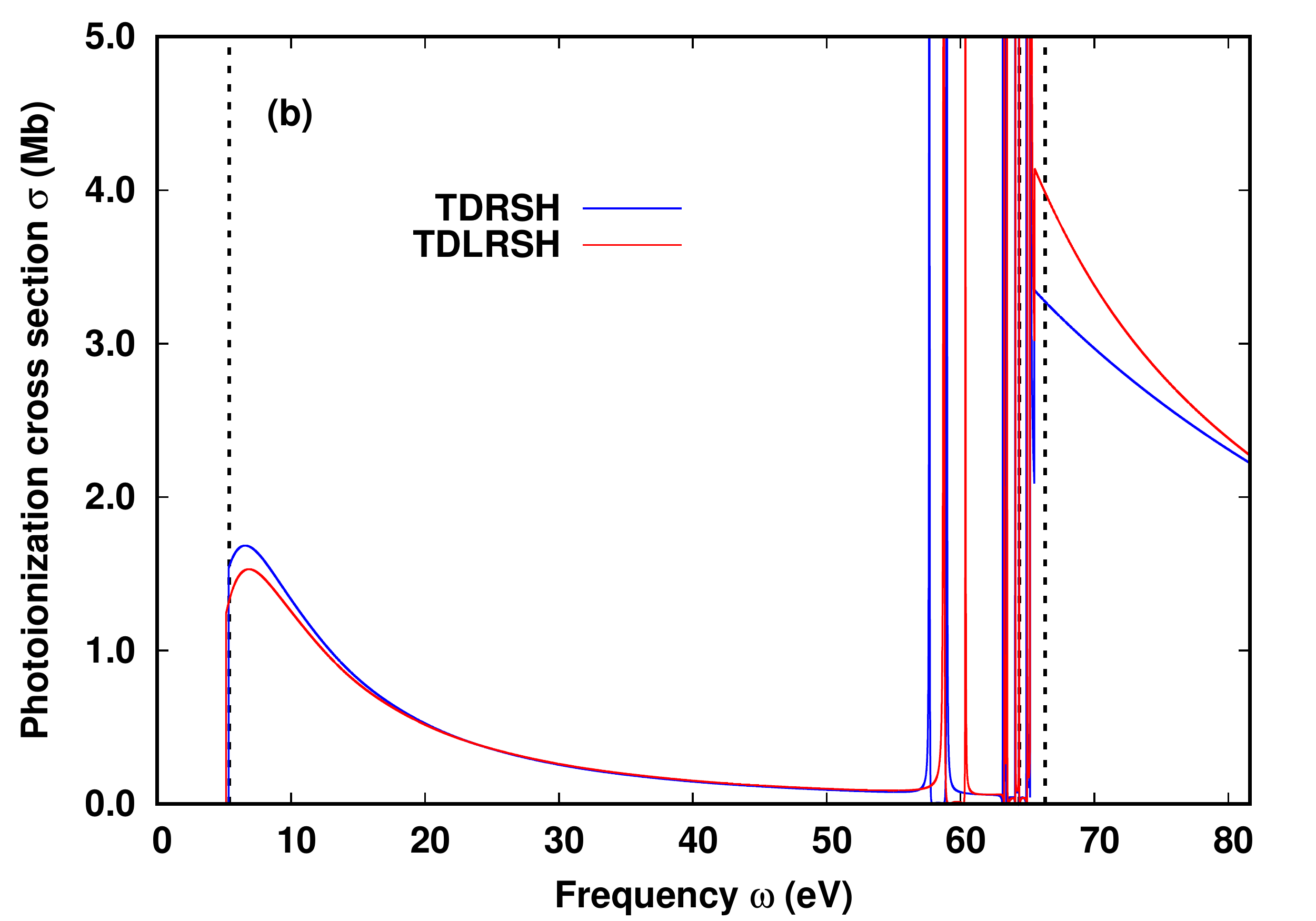}
\caption{Photoionization cross section of the Li atom calculated by (a) TDLDA and TDHF, and by (b) TDRSH and TDLRSH (using the optimal adimensional range-separation parameters determined in Section~\ref{sec:orbenergies}, i.e. $\tilde{\mu}_\text{RSH}=1.431$ for TDRSH and $\tilde{\mu}_\text{LRSH}=0.560$ for TDLRSH). The vertical dashed lines correspond to the experimental IP (5.392 eV)~\cite{NIST-INC-21} and the 1s$_\downarrow$ and 1s$_\uparrow$ ionization edges (64.41 and 66.31 eV, respectively)~\cite{LanVieHemMenWehBec-PRA-91}.}
\label{fig:photoionization}
\end{figure*}

Figure~\ref{fig:photoionization} reports the photoionization cross section calculated by TDLDA, TDHF, TDRSH, and TDLRSH (using the optimal adimensional range-separation parameters determined in Section~\ref{sec:orbenergies}).

The TDLDA photoionization spectrum starts at a too low ionization threshold and the cross section is zero at the threshold, in agreement with the Wigner-threshold law~\cite{Wig-PR-48,SadBohCavEsrFabMacRau-JPB-00} for potentials lacking a long-range attractive $-1/r$ Coulomb tail. At the scale of the plot, the TDLDA 1s$_\uparrow$ and 1s$_\downarrow$ ionization edges are superimposed and occur at a much too low energy. The TDLDA photoionization spectrum contains only the two first 1s$_\uparrow$ $\to$ 2p$_\uparrow$ and 1s$_\downarrow$ $\to$ 2p$_\downarrow$ core resonances, the other core single-excited resonances (involving the orbitals 3p, 4p, etc.) having dissolved into the continuum beyond the 1s ionization edge.

The TDHF photoionization spectrum starts at an ionization threshold very close to the exact value, and the cross section is not zero at the threshold. Again, at the scale of the plot, the TDHF 1s$_\uparrow$ and 1s$_\downarrow$ ionization edges are almost superimposed and occur at a too high energy. In contrast to TDLDA, the TDHF photoionization spectrum contains not only the 1s$_\uparrow$ $\to$ 2p$_\uparrow$ and 1s$_\downarrow$ $\to$ 2p$_\downarrow$ core resonances, but also two intertwined series of single-excited core resonances to Rydberg states (1s$_\uparrow$ $\to$ 3p$_\uparrow$, 1s$_\uparrow$ $\to$ 4p$_\uparrow$, etc., and 1s$_\downarrow$ $\to$ 3p$_\downarrow$, 1s$_\downarrow$ $\to$ 4p$_\downarrow$, etc.) converging toward the 1s$_\uparrow$ and 1s$_\downarrow$ ionization edges, respectively.

The TDRSH and TDLRSH photoionization spectra (using the optimal adimensional range-separation parameters determined in Section~\ref{sec:orbenergies}) display roughly the same features. They both start very close to the exact ionization threshold. For both TDRSH and TDLRSH, the 1s$_\uparrow$ and 1s$_\downarrow$ ionization edges (which are not resolved at the scale of the plot) occur near the experimental ionization edges, as expected since the adimensional range-separation parameter had been adjusted for this purpose. Similar to TDHF, both the TDRSH and TDLRSH photoionization spectra display a series of core resonances. In comparison to TDRSH, TDLRSH gives smaller cross sections in the 2s continuum region (near 5 eV) and larger cross sections in the 1s continuum region (near 70 eV).

\subsection{Core resonances}

\begin{figure*}
\includegraphics[scale=0.3,angle=-90]{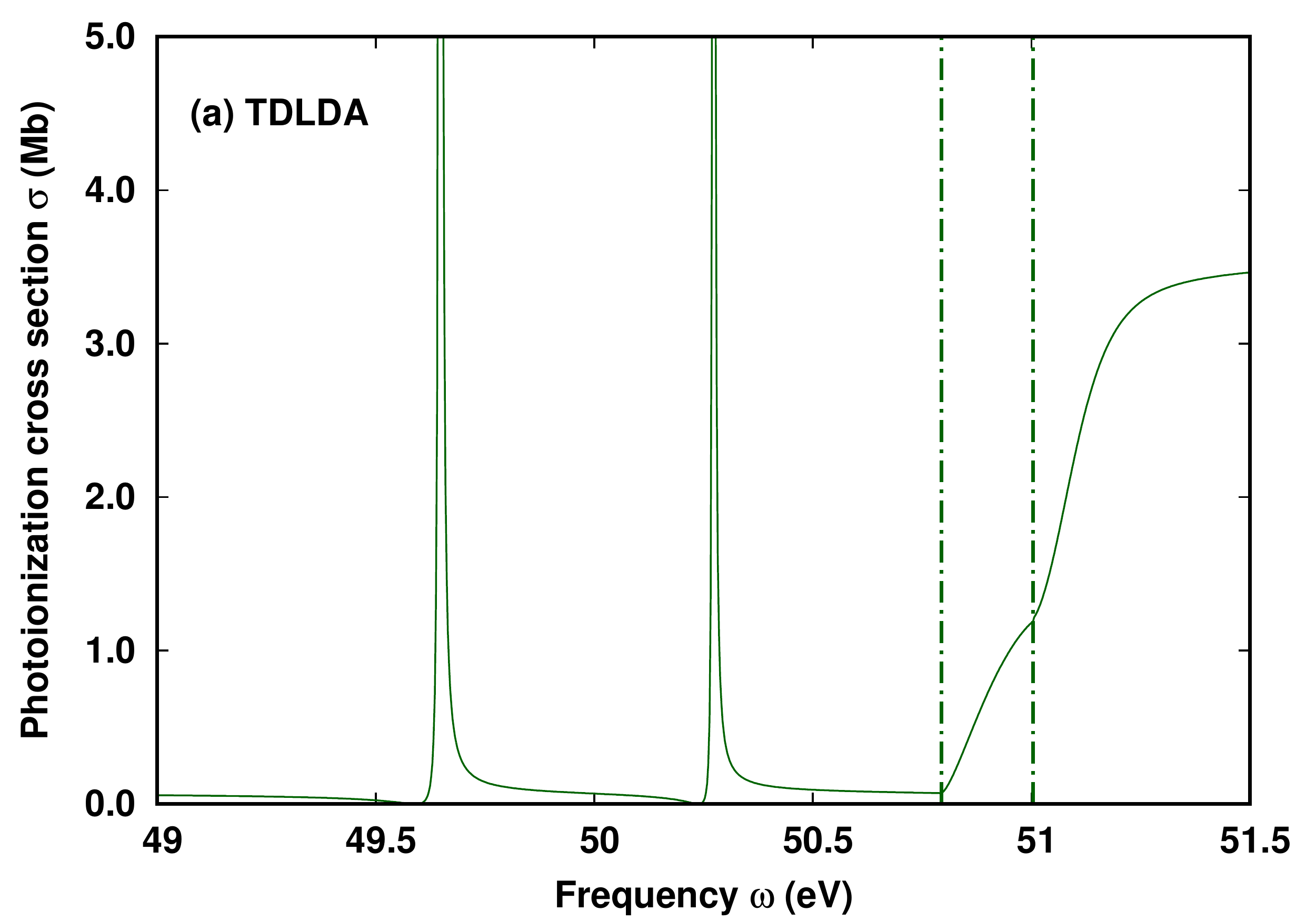}
\includegraphics[scale=0.3,angle=-90]{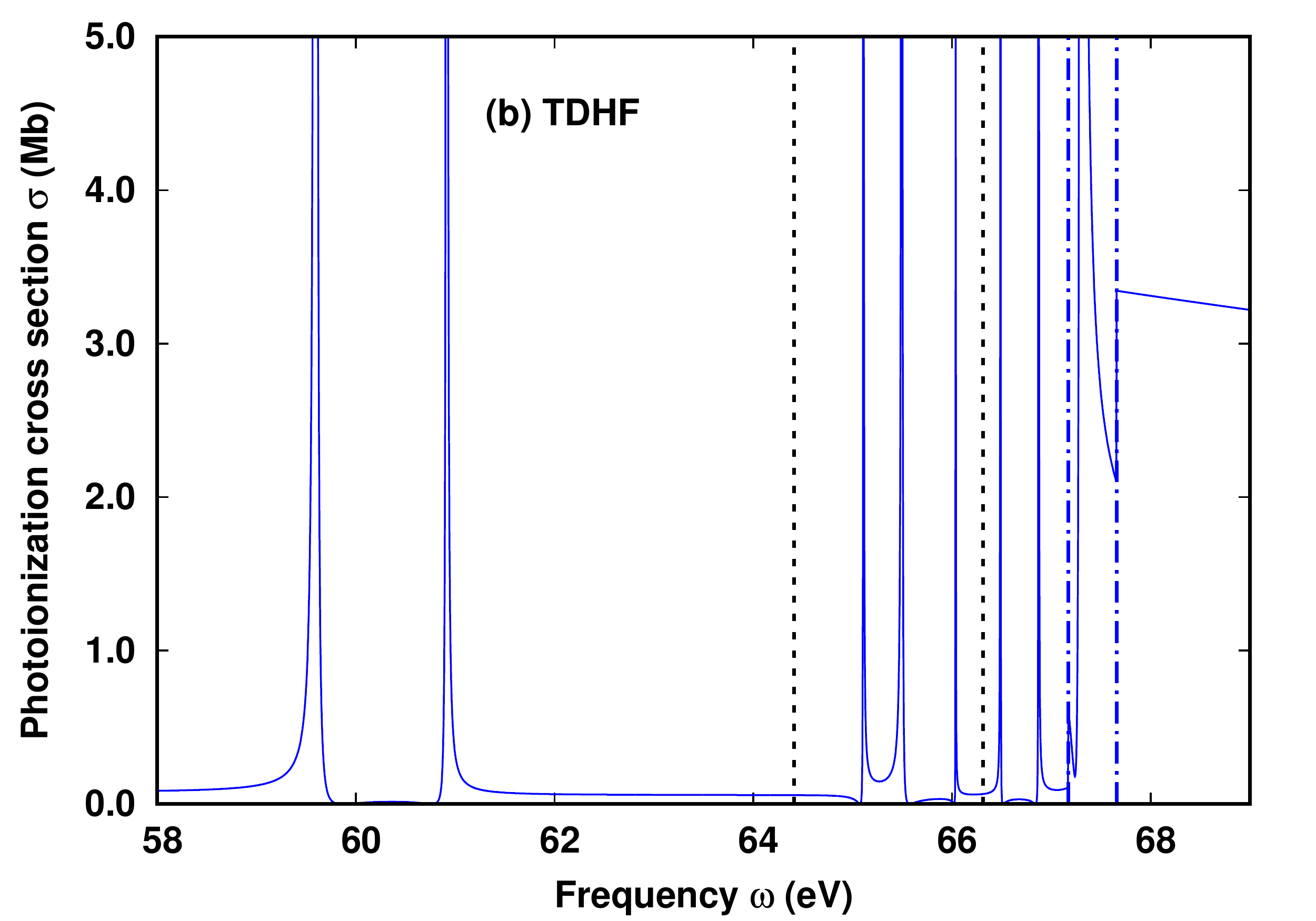}
\includegraphics[scale=0.3,angle=-90]{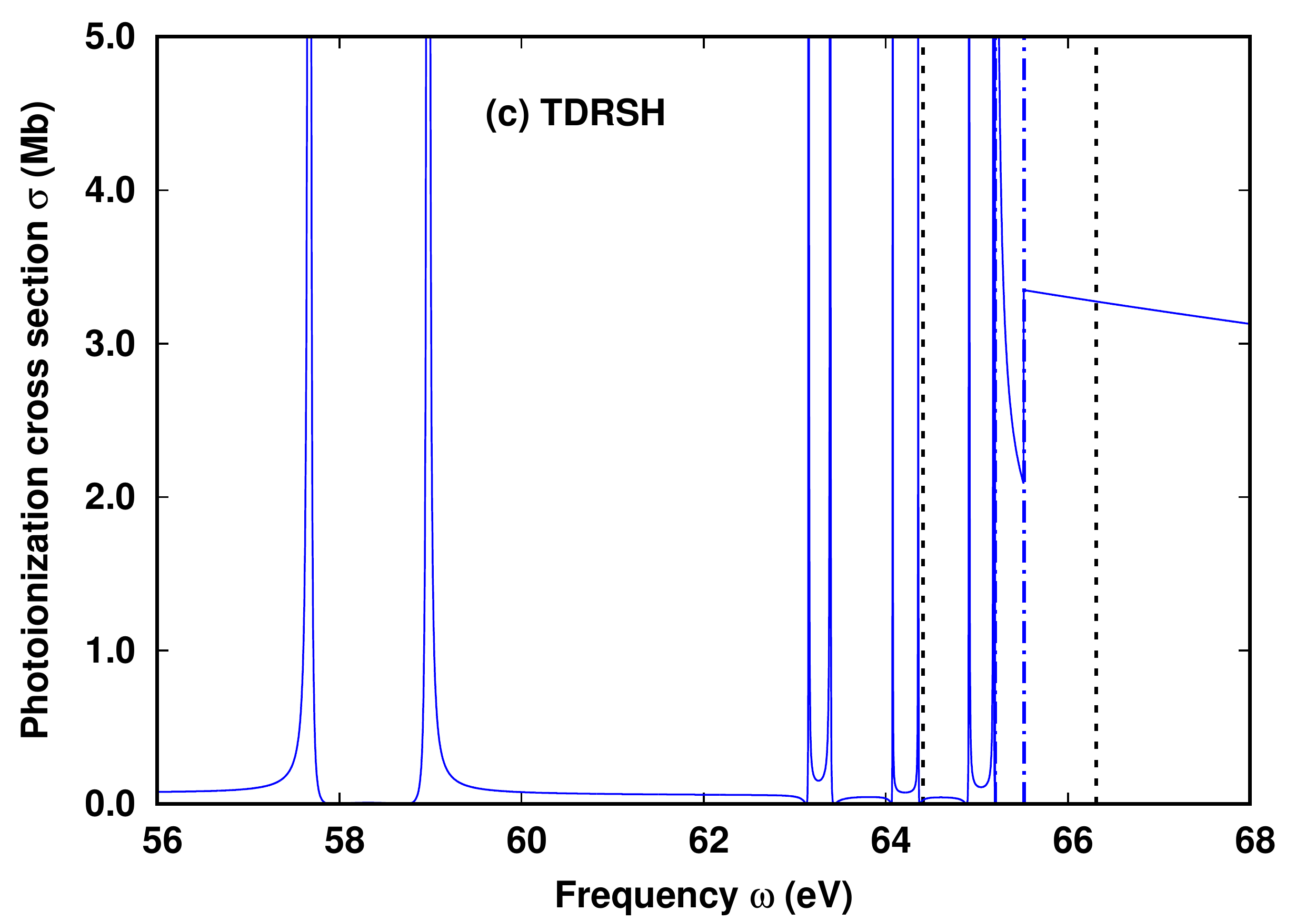}
\includegraphics[scale=0.3,angle=-90]{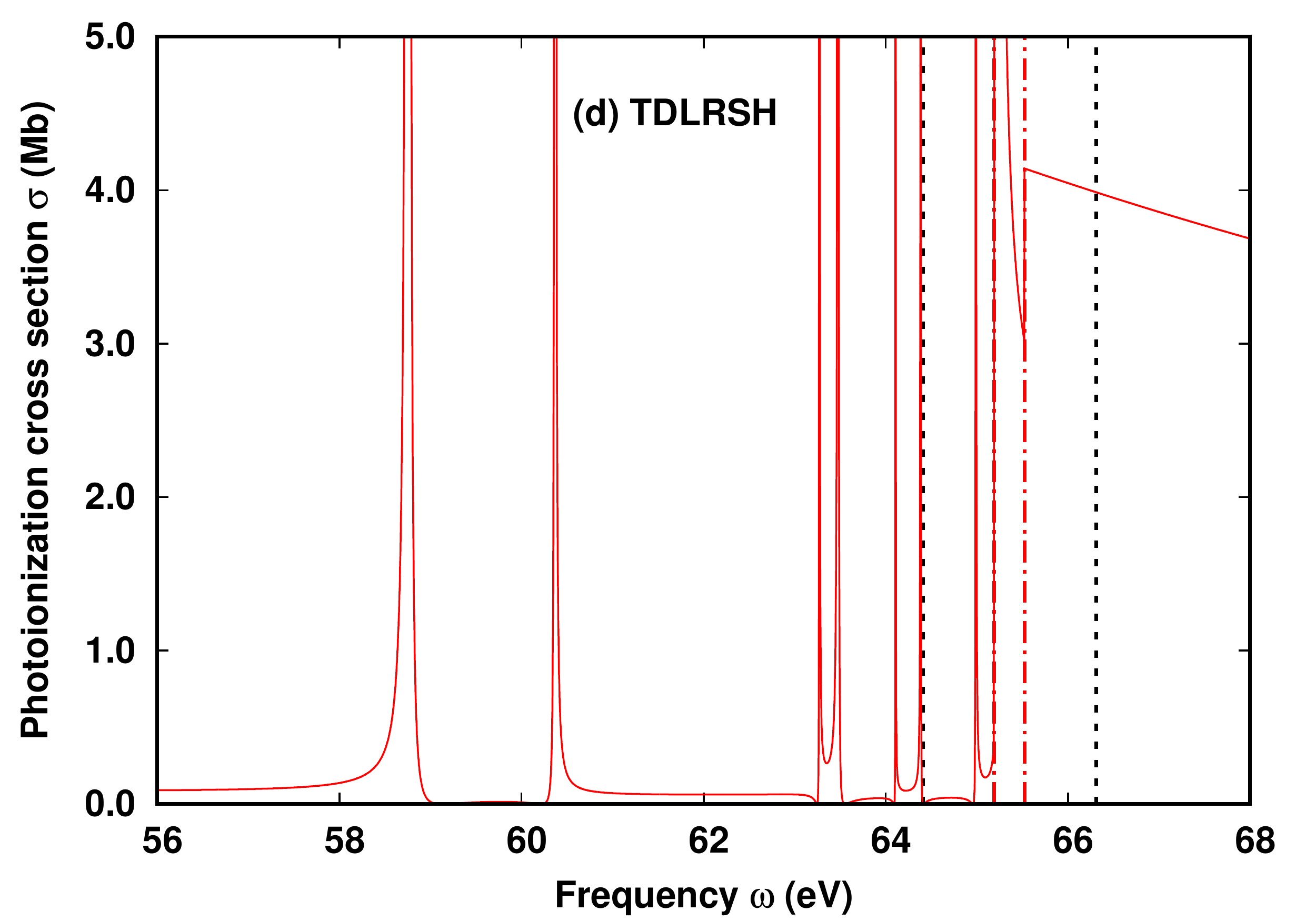}
\caption{Core resonance 1s$_\uparrow$ $\to$ $n$p$_\uparrow$ and 1s$_\downarrow$ $\to$ $n$p$_\downarrow$ of the Li atom calculated by (a) TDLDA, (b) TDHF, (c) TDRSH, and (d) TDLRSH (using the optimal adimensional range-separation parameters determined in Section~\ref{sec:orbenergies}, i.e. $\tilde{\mu}_\text{RSH}=1.431$ for TDRSH and $\tilde{\mu}_\text{LRSH}=0.560$ for TDLRSH). The vertical dash-dotted colored lines correspond the 1s$_\downarrow$ and 1s$_\uparrow$ ionization edges of the method considered, and the vertical dashed black lines correspond to the experimental 1s$_\downarrow$ and 1s$_\uparrow$ ionization edges (64.41 and 66.31 eV, respectively)~\cite{LanVieHemMenWehBec-PRA-91}. 
}
\label{fig:resonance}
\end{figure*}

Figure~\ref{fig:resonance} shows the TDLDA, TDHF, TDRSH, and TDLRSH photoionization spectra in the energy region of the core (Feshbach-type) resonances 1s$_\uparrow$ $\to$ $n$p$_\uparrow$ and 1s$_\downarrow$ $\to$ $n$p$_\downarrow$. In all cases, the cross section follows a characteristic asymmetric Fano lineshape which can be fitted to the analytical expression~\cite{FanCoo-PR-65,SteDecLis-JPB-95}
\begin{equation}
\sigma = \sigma_0 (1+a \epsilon) \left[ \rho^2 \frac{(q+\epsilon)^2}{1+\epsilon^2} -\rho^2+1\right],
\label{sigmafit}
\end{equation}
where $\epsilon = 2({\omega-E_\text{R}})/{\Gamma}$.
Here, $E_\text{R}$ is the resonance energy, $\Gamma$ is the resonance width (or inverse lifetime), $q$ is the asymmetry Fano parameter, $\sigma_0$ is the total background cross section, $a$ is a coefficient for the total background linear drift, and $\rho^2$ is the ratio between the background cross section for transitions to continuum states that interact with the discrete resonant state and the total background cross section. The fitted parameters for the 1s$_\uparrow$ $\to$ 2p$_\uparrow$, 1s$_\downarrow$ $\to$ 2p$_\downarrow$, 1s$_\uparrow$ $\to$ 3p$_\uparrow$, 1s$_\uparrow$ $\to$ 3p$_\uparrow$, and 1s$_\uparrow$ $\to$ 3p$_\uparrow$ resonances are given in Table~\ref{tab:resonance}. For the fitting procedure, the cross section at the resonance energy $\sigma(E_\text{R})$ was included in the data as the asymmetry parameter $q$ is very sensitive to the value of the cross section at the peak. To attribute the correct spin to each resonance line, for each method, we have just calculated photoionization spectra with uncoupled spin-$\uparrow$ and spin-$\downarrow$ excitations (not shown), giving resonances with definite spin which are very close to the original ones.

As references, we have included in Table~\ref{tab:resonance}, the experimental resonance energies~\cite{CanParTonToz-JOSA-77}, as well as accurate results obtained with the R-matrix method~\cite{ZatFro-JPB-00,LiLiuZhaYe-MPLB-16} and the saddle-point complex-rotation (SPCR) method~\cite{Chu-PRA-97,Chu-RPC-04}. The first core resonance cross-section profiles obtained with the R-matrix can be found in Fig. 1a of Ref.~\onlinecite{LiLiuZhaYe-MPLB-16}.

\begin{table*}
\label{tab:resonance}
\caption{Resonance energy $E_\text{R}$, resonance width $\Gamma$, Fano asymmetric parameter $q$, total background cross section $\sigma_0$, background ratio parameter $\rho^2$, background linear drift $a$, and maximum value of the cross section at the resonance energy $\sigma(E_\text{R})$ for the 1s$_\uparrow$ $\to$ 2p$_\uparrow$, 1s$_\downarrow$ $\to$ 2p$_\downarrow$, 1s$_\uparrow$ $\to$ 3p$_\uparrow$, 1s$_\uparrow$ $\to$ 3p$_\uparrow$, and 1s$_\uparrow$ $\to$ 3p$_\uparrow$ core resonances of the Li atom calculated by TDLDA, TDHF, TDRSH, and TDLRSH (using the optimal adimensional range-separation parameters determined in Section~\ref{sec:orbenergies}, i.e. $\tilde{\mu}=1.431$ for TDRSH and $\tilde{\mu}=0.560$ for TDLRSH). 
As references, we also report experimental values and accurate results from R-matrix and SPCR calculations.
}
\begin{tabular}{l c c c c c c c}
\hline\hline
                       & $E_\text{R}$ (eV) & $\Gamma$ (meV) & $q$ & $\sigma_0$  (Mb)  & $\rho^2$ & $a$  & $\sigma(E_\text{R})$ (Mb)\\
\hline\\[-0.3cm]
 resonance 1s$_\uparrow$ $\to$ 2p$_\uparrow$ [configuration 1s$\,$(2s2p)$^3$P]\\[0.1cm]
  TDLDA                      & 49.648  & 0.279  &  403.52  & 0.061  & 1.0072 &  -4.88 $\cdot 10^{-5}$         & 10047.3             \\
  TDHF                       & 59.595  & 5.618  & -93.67   & 0.051  & 1.0448 &  -4.96 $\cdot 10^{-4}$         & 469.2         \\
  TDRSH                      & 57.672  & 2.874  & -170.78  & 0.045  & 1.0074 &  -3.66 $\cdot 10^{-4}$	      & 1326.5               \\
  TDLRSH                     & 58.756  & 5.439  & -136.31  & 0.055  & 1.0288 &  -3.33 $\cdot 10^{-4}$         & 1060.5                \\[0.05cm]
  R-matrix$^a$               & 58.916  &  3.48 \\
  R-matrix$^b$               & 58.898  &  3.99 \\
  SPCR$^c$                   & 58.910  &  3.33 &          &        &        &          &      5164 \\
  Experiment$^d$  	     &  58.909 &    &    &            \\
\\
 resonance 1s$_\downarrow$ $\to$ 2p$_\downarrow$ [configuration 1s$\,$(2s2p)$^1$P]\\[0.1cm]
  TDLDA                      & 50.273  & 0.142  & 488.14   & 0.076  & 1      & -2.76 $\cdot 10^{-5}$       & 18162.4              \\
  TDHF                       & 60.915  & 0.174  & 1692.39  & 0.042  & 1.0054 &  1.86 $\cdot 10^{-5}$       & 121319.2             \\
  TDRSH                      & 58.974  & 0.566  & 891.62   & 0.042  & 1.0443 &  5.83 $\cdot 10^{-5}$       & 34676.3             \\
  TDLRSH                     & 60.370  & 0.273  & 1132.04  & 0.039  & 1.0160 &  4.12 $\cdot 10^{-5}$       & 50323.1              \\[0.05cm]
  R-matrix$^a$               & 60.409  &  9.54  \\
  R-matrix$^b$               & 60.357  & 10.52  \\
  SPCR$^c$                   & 60.398  &   9.56 &          &        &        &                         &          84.3  \\
  Experiment$^d$  	     &  60.392 &    &    &            \\
\\
 resonance 1s$_\downarrow$ $\to$ 3p$_\downarrow$ [configuration (1s2s)$^3$S 3p]\\[0.1cm]
  TDHF                       & 65.109  & 0.454  & 149.05   & 0.083  & 1      & 1.52 $\cdot 10^{-4}$	      & 1824.0               \\
  TDRSH                      & 63.155  & 0.410  & 128.83   & 0.077  & 1      & 4.15 $\cdot 10^{-4}$	      & 1268.7               \\
  TDLRSH                     & 63.272  & 0.675  & 79.74    & 0.106  & 1      & 1.33 $\cdot 10^{-3}$           & 671.8                 \\[0.05cm]
  R-matrix$^a$               & 62.423  & 0.196  \\
  R-matrix$^b$               & 62.415  & 0.214  \\
  SPCR$^c$                   & 62.417  &  0.203 &          &        &        &                        &     14630 \\
  Experiment$^d$  	     &  62.417 &    &    &            \\
\\
 resonance 1s$_\uparrow$ $\to$ 3p$_\uparrow$ [configuration (1s2s)$^1$S 3p]\\[0.1cm]
  TDHF                       & 65.495  & 0.580  & -276.35  & 0.062  & 1      & -2.54 $\cdot 10^{-4}$	      & 4741.2               \\
  TDRSH                      & 63.391  & 0.156  & -546.66  & 0.066  & 1      & -1.23 $\cdot 10^{-4}$	      & 19821.8              \\
  TDLRSH                     & 63.476  & 0.488  & -683.20  & 0.016  & 1      & -3.58 $\cdot 10^{-3}$              & 7372.2                \\[0.05cm]
  R-matrix$^a$               & 64.051  & 0.352\\
  SPCR$^c$                   & 64.050  &  0.391 &          &        &        &                        &     173   \\
  Experiment$^d$  	     & 64.052  &    &    &            \\
  \hline\hline
\multicolumn{5}{l}{$^a$From Ref.~\onlinecite{ZatFro-JPB-00}.}\\
\multicolumn{5}{l}{$^b$From Ref.~\onlinecite{LiLiuZhaYe-MPLB-16}.}\\
\multicolumn{5}{l}{$^c$From Ref.~\onlinecite{Chu-PRA-97} (see also Ref.~\onlinecite{Chu-RPC-04}).}\\
\multicolumn{5}{l}{$^d$From Ref.~\onlinecite{CanParTonToz-JOSA-77}.}
\end{tabular}
\end{table*}

The TDLDA 1s$_\uparrow$ $\to$ 2p$_\uparrow$ and 1s$_\downarrow$ $\to$ 2p$_\downarrow$ resonances occur at much too low energies (by 9.3 eV and 10.1 eV, respectively). With TDHF, the 1s$_\uparrow$ $\to$ 2p$_\uparrow$ and 1s$_\downarrow$ $\to$ 2p$_\downarrow$ resonances have slightly too high energies (by 0.7 eV and 0.5 eV, respectively) and the errors increase for the 1s$_\downarrow$ $\to$ 3p$_\downarrow$ and 1s$_\uparrow$ $\to$ 3p$_\uparrow$ resonances (with energies overestimated by 2.7 eV and 1.4 eV, respectively). TDRSH does not systematically improve over TDHF: although TDRSH gives 1s$_\downarrow$ $\to$ 3p$_\downarrow$ and 1s$_\uparrow$ $\to$ 3p$_\uparrow$ resonance energies with much smaller absolute errors (0.7 eV for both resonances) compared to TDHF, it gives 1s$_\uparrow$ $\to$ 2p$_\uparrow$ and 1s$_\downarrow$ $\to$ 2p$_\downarrow$ resonance energies with larger absolute errors (1.2 eV and 1.4 eV, respectively). By contrast, TDLRSH provides a systematic improvement over TDHF: it gives 1s$_\uparrow$ $\to$ 2p$_\uparrow$ and 1s$_\downarrow$ $\to$ 2p$_\downarrow$ resonance energies with absolute errors of 0.15 eV and 0.02 eV, respectively, and 1s$_\downarrow$ $\to$ 3p$_\downarrow$ and 1s$_\uparrow$ $\to$ 3p$_\uparrow$ resonance energies with absolute errors of 0.9 eV and 0.6 eV, respectively.

The resonance widths $\Gamma$ and Fano asymmetric parameters $q$, which determined the shape of the resonances, are very sensitive to the method employed. The resonance widths $\Gamma$ should correspond to the decay rate of the core resonances through the Auger process 1s2s$n$p $\to$ 1s$^2$ + e (for $n=2$ or $3$). Since the last configuration is a single excitation with respect to the ground-state configuration, one can a priori hope to obtain reasonable resonance widths with the present adiabatic TDDFT/TDHF-type methods (in contrast with the situation of the core resonances of the Be atom with Auger decays involving a double excitation~\cite{SchZapLevCanLupTou-JCP-22}). While TDLDA turns out to give much too small widths $\Gamma$ (by one or two orders of magnitude) for the first two core resonances, TDHF gives indeed reasonable widths $\Gamma$ (of the correct order of magnitude) for the 1s$_\uparrow$ $\to$ 2p$_\uparrow$, 1s$_\downarrow$ $\to$ 3p$_\downarrow$, and 1s$_\uparrow$ $\to$ 3p$_\uparrow$ resonances. TDHF only largely underestimates (by about a factor $50$) the width of the 1s$_\downarrow$ $\to$ 2p$_\downarrow$ resonance. According to the reference methods, this last resonance has much larger decay rate than the other resonances considered here, with could be explained by a larger proximity of the two electrons involved in the Auger process in the 1s$\,$(2s2p)$^1$P resonance state~\cite{Chu-PRA-99}. This particular feature of the 1s$\,$(2s2p)$^1$P resonance state is not reproduced by TDHF. As regards now TDRSH and TDLRSH, they give resonance widths of the same order of magnitude than the TDHF ones, sometimes smaller and sometimes larger, without any clear pattern emerging.

\section{Conclusion}
\label{sec:conclusion}

In this work, we have considered the calculation of photoionization spectra of open-shell systems using two variants of range-separated TDDFT, namely, TDRSH, which uses a global range-separation parameter, and TDLRSH, which uses a local range-separation parameter, and compared with standard TDLDA and TDHF. For this, we have formulated a spin-unrestricted linear-response Sternheimer approach in a non-orthogonal B-spline basis set and using appropriate frequency-dependent boundary conditions. We have illustrated this approach on the photoionization spectrum of the Li atom, focusing in particular on the core resonances.
 
TDRSH and TDLRSH provide a big improvement over TDLDA and a small improvement over TDHF. Moreover, TDLRSH tends to provide slightly more accurate resonance energies than TDRSH. This suggests that TDRSH and TDLRSH are adequate simple methods for estimating single-electron photoionization spectra of open-shell systems, even though neither TDRSH or TDLRSH can compete with more accurate methods such as the R-matrix and SPCR methods, especially for the calculation of resonance widths. 

To extend this work to general molecular systems, the present linear-response Sternheimer approach could be implemented with Gaussian basis sets and should be extended from spherical boundary conditions to general nonlocal Robin boundary conditions. To improve the accuracy, the present approach could be extended to range-separated multiconfiguration TDDFT~\cite{FroKneJen-JCP-13}.

\section*{Acknowledgements}
This project has received funding from the CNRS Emergence@INC2021 program (project OPTLHYB) and from the European Research Council (ERC) under the European Union's Horizon 2020 research and innovation programme Grant agreement No. 810367 (EMC2).

\section*{Author Declarations}
The authors have no conflicts to disclose.

\section*{Data Availability}
The data that support the findings of this study are available from the corresponding author upon reasonable request.


\end{document}